# Aminated TiO$_2$ nanotube as a Photoelectrochemical Water Splitting photoanode

S. Hejazi[1], N.T. Nguyen[1], A. Mazare[1], P. Schmuki[1,2,*]

[1] Department of Materials Science, WW4-LKO, University of Erlangen-Nuremberg, Martensstrasse 7, D-91058 Erlangen, Germany
[2] Chemistry Department, Faculty of Sciences, King Abdul-Aziz University, 80203 Jeddah, Saudi Arabia

*Corresponding author: Prof. Patrik Schmuki, schmuki@ww.uni-erlangen.de

**Abstract**

The present work reports on the enhancement of TiO$_2$ nanotubes photoelectrochemical water splitting rate by decorating the nanostructure with an amine layer in a hydrothermal process using diethylenetriamine (DETA). The aminate coated TiO$_2$ tubes show a stable improvement of the photoresponse in both UV and visible light spectrum and under hydrothermal conditions, 4-fold increase of the photoelectrochemical water splitting rate is observed. From intensity modulated photocurrent spectroscopy (IMPS) measurements significantly faster electron transport times are observed indicating a surface passivating effect of the N-decoration.

1. **Introduction**

Since the first report by Fujishima and Honda [1] on the photoelectrolysis of water into $H_2$ and $O_2$ on a $TiO_2$ electrode, intensive research has been devoted to hydrogen production from water on $TiO_2$ photoanodes [2–6]. In photoelectrochemical water splitting on a semiconductor photoanode, conduction band electrons are transferred to a counter electrode (typically Pt) to reduce water to $H_2$ and valence band holes are used to oxidize water [7]. There are three key factors influencing the photoelectrolysis reaction: *(1)* light absorption and carrier excitation (electron-hole formation), *(2)* charge separation and *(3)* charge transfer [7]. $TiO_2$ remains still one of the most investigated photocatalytic material, as it is stable against photocorrosion, cost efficient and can be nanostructured.

One-dimensional $TiO_2$ nanoarchitectures, as nanotubes or nanowires, grant unidirectional pathways for photoexicited charge carriers and can diminish electron-hole recombination pathways [8,9]. $TiO_2$ being a wide band gap semiconductor (3-3.2 eV) is only able to produce photocurrent in the UV range, and only able to make use of less than 5 percent of the solar spectrum [1,2,6,9]. To improve the visible light response, extensive research was dedicated to band-gap engineering (doping) of $TiO_2$ using a wide range of transition metals such as V, Ce, Mn and N, or nonmetals as C, N and S [10–19] – for an overview see refs [20].

From the available dopant materials, N-doping is used for varying the optical and electrical properties of $TiO_2$, and can be achieved by various techniques such as powder metallurgy, wet processes or annealing in ammonia media [21–26]. At low to medium nitrogen concentrations in the $TiO_2$ lattice, most reports describe the formation of N-substitutional states close to the valence band of $TiO_2$ [21–25], successfully narrowing the optical absorption edge and causing the well-established activation of $TiO_2$ in the visible range [23–25]. From the variety of doping methods, annealing of $TiO_2$ nanostructures in $NH_3$ atmosphere is one of the simplest methods for N doping of anodic $TiO_2$ nanotubes [27,28],

Despite the advantages of nitrogen doping, increasing the nitrogen content in the $TiO_2$ lattice usually causes a decrease of the overall quantum efficiency due to increasing electron-hole recombination site [29,30].

Literature data already shows that amines can be adsorbed on $TiO_2$ [31], and that this adsorbed nitrogen species affect the visible light photoresponse of $TiO_2$ powders [32–34], but there is hardly any information regarding the interaction of amines decorated on nanotubular $TiO_2$ and their effect on the photoelectrochemical response.

In the present study we report the synthesis of aminated $TiO_2$ nanotube by a hydrothermal treatment of anodic $TiO_2$ nanotubes in diethylenetriamine. The quantity of the adsorbed amine depends on the temperature and time of the hydrothermal, and for these aminated structures an increase of the photoresponse in both UV and visible range of light spectrum is observed.

## 2. Experimental

### 2.1 Electrochemical anodization

Pure titanium (Ti) foils were cleaned by sonication in acetone, ethanol and deionized water, followed by drying in $N_2$ gas flow. The $TiO_2$ nanotubes were formed by anodizing titanium foils in ethylene glycol electrolyte containing $NH_4F$ 0.15 M and 3 wt% $H_2O$, at 60 V for 17 min. The DC potential was applied by using a power supply. Right after the anodization, the samples were immersed in ethanol for 3 hour (to remove the remnants from the organic electrolyte), and then dried under a $N_2$ gas flow. Then, the nanotubes were annealed at 450 °C in air for 1 h using a Rapid Thermal Annealer, with a heating and cooling rate of 30 °C $min^{-1}$.

### 2.2 Hydrothermal treatment of the $TiO_2$ nanotubes

After annealing, the nanotubular samples were immersed into an autoclave (vol. 200 ml) containing pure diethylenetriamine (SigmaAldrich; vol. 100 ml), and then heated in an oven at different temperatures (e.g. 150°C, 200°C or 250°C) for 4h up to 16h. Afterwards, samples were washed by distilled water (DI) water and dried in a nitrogen flow.

2.3 Morphological, structural and chemical characterization

A field-emission scanning electron microscope (FE-SEM, Hitachi S4800) was used to characterize the morphology of the nanotubular samples. The crystallinity of the samples was analyzed by X-ray diffraction (XRD) performed with a X′pert Philips MPD (equipped with a Panalytical X'celerator detector and using graphite monochromized Cu Kα radiation ($\lambda$ = 1.540 56 Å)).

The composition and the chemical state of the bare and aminated $TiO_2$ nanotubes were characterized using X-ray photoelectron spectroscopy (XPS, PHI 5600, US), and the spectra were shifted in relation to the Ti2p signal at 458.2 eV (the N1s, C1s and O1s peaks were fitted with the Multipak software).

In order to further investigate the amine presence of $TiO_2$ nanostructures, time-of-flight secondary ion mass spectrometry (TOF-SIMS) surface spectra in positive and negative polarity were recorded on a TOF SIMS 5 instrument (ION-TOF, Münster, Germany). Signals were identified according to their isotopic pattern as well as exact mass. Spectra were calibrated to $CH_2^-$, $C_2^-$, $CN^-$ and $CNO^-$ (negative polarity) and $C^+$, $CH^+$, $CH_2^+$, $CH_3^+$ and $C_7H_7^+$ (positive polarity) and Poisson correction was employed.

2.4 Incident photon to current conversion efficiency (IPCE)

Photoelectrochemical characterization was carried out with a setup consisting of a 150 W Xe arc lamp (LOT-Oriel Instruments) as the irradiation source and a Cornerstone motorized 1/8 m monochromator. Photocurrent spectra were acquired in 0.1 M ($Na_2SO_4$) at a potential of 500 mV (vs. Ag/AgCl). IPCE for each wavelength was calculated according to (Eq. (1)) [35]:

$$IPCE(\%) = \frac{I_{Ph} \times h\upsilon}{p \times q} \times 100 \qquad (1)$$

where $I_{ph}$ is photocurrent density, $P$ is the power density of light, $h\nu$ (≈1240) photon energy of the incident light, $q$ charge of the electron.

2.5 Photoelectrochemical water splitting

The photoelectrochemical water splitting experiments were performed under simulated AM 1.5G illumination supported by a solar simulator, in 1M KOH aqueous solution. A three-electrode configuration was used in the measurement, with the $TiO_2$ nanotubes (before – bare, and after hydrothermal treatment – amination) as a working electrode (photoanode), Ag/AgCl as a reference electrode, and a platinum foil as a counter electrode. Photocurrent versus voltage curves were acquired by scanning the potential from −0.5 to 0.7 with a scan rate of 1 mV.s$^{-1}$ using a Jaissle IMP 88 PC potentiostat. The stability measurements were performed at 0.5 V (vs. Ag/AgCl) in 1 M KOH for 1 hour. Photocurrent transient of bare and aminated $TiO_2$ nanotube were measured under monochromatic (474 nm) laser illumination.

2.6 Intensity modulated photocurrent spectroscopy (IMPS)

Intensity modulated photocurrent spectroscopy (IMPS) measurements were carried out using a Zahner IM6 (Zahner Elektrik, Kronach, Germany) with an UV and Visible modulated light (λ = 369 nm & 475 nm). The photoelectrochemical performance of the samples was analyzed in pure 0.1 M $Na_2SO_4$ solution in a three-electrode configuration, consisting of $TiO_2$ nanotubes as a photoanode, platinum wire as a counter electrode and an Ag/AgCl (3 M NaCl) electrode as a reference.

## 3. Results and Discussion

Previous reports [16,36] have indicated that for photoelectrochemical water-splitting of anodic nanotubular layers, there is an optimal thickness of the nanotubular layers allowing full-light absorption and minimized electron-hole recombination of ≈ 7 μm. Hence, this length range of the $TiO_2$ nanotubes was used in the present work. After anodization, the nanotubular samples were annealed at 450°C in order to convert the amorphous structure to a crystalline one (anatase). The obtained crystalline $TiO_2$ nanotubes (NTs) are shown in Figure 1 a-c and the nanotubes present a typical morphology for NTs obtained in organic electrolytes [8,37], i.e. well-defined nanotubular layers with a tube diameter of ≈ 100nm, a length of ≈ 6.5 μm smooth and ripple free nanotube walls and a double-walled structure.

It is known that for compact or nanoparticulate $TiO_2$[31,33,38], amines can adsorb and influence the visible light photoresponse. For the synthesis of aminated nanoparticulate $TiO_2$, amines are frequently added directly in the hydrothermal synthesis. However, the hydrothermal treatment of $TiO_2$ nanotubes is more delicate (e.g. decay of the nanotubular structure), therefore it is not surprising that amine treatments on $TiO_2$ nanotubes have not been reported. Herein, the amine layer was coated on the $TiO_2$ NTs by a hydrothermal treatment performed in diethylenetriamine (DETA) and it was carried out at different temperatures (150°C, 200°C or 250°C) for 12h. Diethylenetriamine ($C_4H_{13}N_3$) was used, as it presents a higher nitrogen to carbon ratio (0.75) compared to other amines, e.g. dimethylamine (($CH_3$)$_2NH$), ethylamine ($C_2H_7N$) etc.

For all aminated $TiO_2$ NTs samples, we observed no difference in the morphology of the nanotubes compared to the bare $TiO_2$ nanotubes, e.g. comparative SEM images of the bare and aminated NTs (at 250°C) are shown in Figure 1 (d-f vs a-c) – data not shown for 150°C and 200°C. There is no visual or morphological alteration apparent from SEM images that would indicate any film formation on the nanotubes or on the tube tops after the hydrothermal treatment in DETA.

Moreover, the hydrothermal treatment had no influence on the crystallinity of the NTs layers, as observed by XRD i.e. no significant difference of patterns between bare NTs (anatase $TiO_2$ nanotubes – Figure 1.g) and the aminated NTs (aminated at 250°C – Figure 1.h) could be seen (similar XRD data were obtained for 150°C and 200°C, data not shown).

To confirm the presence of the DETA amine on the nanotubular samples after the hydrothermal treatment, high resolution XPS spectra of N1s, C1s, Ti2p and O1s were performed and as seen from Figure 2, an increase in the N1s and C1s peaks is observed, while no significant difference is observed for Ti2p and O1s (data not shown for O1s at 530 eV). First of all, the presence of the nitrogen is clearly detected for all aminated samples (150°C, 200°C and 250°C) from the N1s *spectra*, i.e the N1s peak at ≈ 399.8 eV is evident for all aminated samples and results in higher N at.%, compared to the bare $TiO_2$ nanotubes (bare NTs – with N-pick up from environment), see Figure 2.b and Table 1. In the C1s *spectra*, an increase in the C peaks is observed for all aminated NTs (Figure 2.b and Table 1), due to the presence of C-N and C-C bonds in the DETA [31]. Moreover, for all aminated samples we observe a shift of the C1s peak to ≈286.2 eV, which is due to the C-N bonds (≈286 eV); also, for the 250°C aminated NTs, there is a shoulder at higher binding energies (288-290 eV) which can be ascribed to the presence of both amide carbon (N-C=O, ≈287.8 eV) or carbon in carbonates (O=C-OH, ≈288.5 eV). To emphasize the effect of the thermal amination, we also characterized NTs that were immersed at room temperature in DETA (adsorbed RT): in this case only very little amine is adsorbed on the $TiO_2$ surface, as there are no significant difference in the N at.% compared to the bare NTs, however in the C1s spectra we observe a small peak corresponding to C-N bonds at ≈286.2 eV.

Additionally, in order to distinguish between the chemical states of the nitrogen in the aminated samples, the N1s peak was deconvoluted (e.g. 250°C aminated NTs in Figure 2.d) and the fitting resulted into three peaks at 398.6 eV – assigned to free amines groups, 399.6 eV – to amine groups bonded to the $TiO_2$ surface, and at 401.5 eV corresponding to protonated amines or oxidized N species [39,40]. The deconvolution results for all aminated samples are also listed in Table 1; it is worth pointing out that with increasing the temperature of the hydrothermal treatment there is an increase in the amount of nitrogen and also in the amount of N bonded to the $TiO_2$ surface (after deconvolution). Previous work by Farfan-Arribas et al. [31] reported that generally the adsorption of amines occurs through the formation of a N-Ti bond with the surface $Ti^{4+}$ cations and that on defective surfaces, adsorption occurs on the vacancies as well as the cations.

Selected signals from ToF-SIMS surface profiles are presented in Figure 3 and these correspond to the following selected characteristic fragments in positive polarity: $TiO^+$ (mass 63.94), $C_2H_6N^+$ (mass 44.05), $C_3H_7N_2O_2^+$ (mass 103.05) and while no difference in observed in the fragments corresponding to the $TiO_2$ nanotubes (i.e. $TiO^+$, and also for $Ti^+$), an increase is detected for fragments of the DETA (e.g. $C_2H_6N^+$, $C_3H_7N_2O_2^+$ and also for other fragments such as $C_2H_2N^+$, $C_3H_6NO^+$, $C_2H_7N_2^+$ for which data is not shown here). Moreover, in negative polarity an increase is observed also in fragments such as $CN^-$ (mass 26.00) or $CHN^-$ (mass 27.00). ToF-SIMS surface profile data are correlating with the XPS data, confirming the presence of amine groups on the $TiO_2$ nanotubes.

The photocurrent spectra of differently treated NTs layers are shown in Figure 4a, i.e. bare nanotube layers (as-grown and converted to anatase), hydrothermally treated NTs in DETA (aminated layers at 150°C, 200°C or 250°C) and for DETA adsorbed NTs at room temperature (25°C). The highest photocurrent density is achieved for aminated nanotubes at 250°C, followed by the amination at 200°C and 150°C, respectively. DETA adsorbed at room temperature on the $TiO_2$ NTs does not induce any change in the photocurrent spectra, in addition XPS data indicated no significant amine adsorption compared to the bare NTs, in these conditions. From Figure 4a one can see that amination (i.e. hydrothermal treatment in DETA) increases not only the visible photo response but also the photocurrent response in the UV region. In previously published N-doping research [30,41], N-doping led to a significant decrease in quantum efficiency at the UV region of the solar spectrum and further increasing the nitrogen content caused a decrease of quantum efficiency in both UV and visible region of solar spectrum, as a result of increasing electron-hole recombination sites (traps) [29]. The incident photon to current efficiency (IPCE) reaches values of 7-19% at the UV spectrum maximum and values of 0-3.5% at the visible spectrum (data not shown) supporting that the aminated nanotubes show an increased quantum efficiency in both UV and visible range (due to increasing amount of adsorbed nitrogen species).

Figure 4b presents the $(I_{\text{ph norm}} h\nu)^{0.5}$ vs. $h\nu$ plots, from which the band gap ($E_g$) values can be evaluated. The photoresponse for untreated $TiO_2$ nanotubes (i.e. non-doped $TiO_2$ nanotubes in the form of anatase) is in line with previous work[35] and shows the expected $E_g$ values for anatase at 3.15 eV. However, the significant increase in the magnitude of the photocurrent in the visible region can be related to formation of sub-band gaps with much lower values, i.e. with the value of $\approx$ 2.2eV for the 250°C aminated $TiO_2$ NTs. Hence, this can be due to the adsorption of nitrogen containing species on the surface of $TiO_2$ nanotubes that could result in a sub band gap of N2p between O2p and Ti3d state[29,42].

The duration of the hydrothermal treatment was optimized in regard of obtaining a maximum photocurrent. Namely, various times were investigated starting from 6h up to 18h (see Figure 4c) and at 12h a plateau of maximum photocurrent was reached – thus, the time of 12h was considered optimal for the hydrothermal treatment.

Intensity modulated photocurrent spectra (IMPS) for the bare and 250 ºC aminated nanotubes are shown in Figure 4d for monochromatic 396 and 475 nm light illumination, and faster electron transport times are observed for the aminated NTs. Results under UV illumination clearly show that essentially the same power-law dependence on the incident light intensity and lower transport time is in aminated sample. Such power-law dependence is normally attributed to the nature of the trap distribution [43–45]. Recent studies [46] show that the traps limiting transport in $TiO_2$ nanoparticle films are located predominantly on the surface of the oxide. Hence, the shift in transport time occurs due to the decreasing of trap sites at the surface and one can say that the hydrothermal treatment can decrease the number of traps on the surface and thus causing the increase in the UV response, in the IPCE spectrum. The difference between the transport times coefficient can also indicate that the aminated NTs show a power-law dependence, however a different intensity does not such a definite impact in the case of the bare anatase $TiO_2$ nanotubes.

The aminated $TiO_2$ nanotubes were further evaluated for photoelectrochemical water splitting under simulated sunlight conditions (AM 1.5 at 100mWcm$^{-2}$) – see Figure 5a. From the transient-photocurrent versus potential curves, it is evident that with amination of the $TiO_2$ nanotubular samples the photocurrent density is significantly enhanced. Furthermore, there is a correlation between the temperature of the hydrothermal treatment, namely the temperature at which the amination is performed, and the increase in the photocurrent density. We observed that a higher temperature of the hydrothermal treatment results in increasing adsorbed nitrogen species on the $TiO_2$ surface that results in an improvement of their photoelectrochemical properties. The photocurrent density of the 250 ºC aminated $TiO_2$ NTs presents a 4-fold increase compared to the bare $TiO_2$ nanotubes (anatase). In addition, the aminated layers present a good photostability over 1h – see Figure 5b.

The photocurrent transients of bare and 250 ºC aminated $TiO_2$ nanotube and under monochromatic laser ($\lambda$ = 474 nm) are shown in Figure 5c and confirm the photocurrent response of aminated NTs in visible light and further corroborate the IPCE measurement (Figure 4). However, the photocurrent transient registered for the bare NTs has negligible values and can be due to the presence of remnant carbon species in the nanotubes (i.e. from the carbon contaminated inner layer of the classical double-walled nanotubular structure) [8,42]; generally, this absorption feature is of a broadband nature, namely it extends up to ca. 600 nm [47]. Recent work [47] points out that such a small current response for anatase $TiO_2$ nanotubes can be attributed to a photo sensitization of the carbon contamination in the nanotubes.

Finally, it is noteworthy that significance increase in photoelectrochemical water splitting of the nanotubular structures is achieved by hydrothermal treatment of anodic $TiO_2$ nanotubes in amines, and it is plausible that such treatments can be transferred to other state-of-the art nanotubular structures (e.g. and even to most recent generation of single-walled or $TiO_2$ nanoparticle decorated nanotubes) or to other $TiO_2$ nanostructures.

**Conclusions**

In summary, we have shown that aminated $TiO_2$ nanotubular structures (via hydrothermal treatment of anodic $TiO_2$ nanotubes in diethylenediamine) represent a straightforward method for increasing the photoelectrochemical response of nanotubes. The amination of the nanotubular structure does not modify its morphology or crystal structure; however, it leads to a decoration of the tube with an amine layer. This leads to an increase in the photocurrents for both UV and visible regions of the light spectrum. Compared to bare $TiO_2$ nanotubes, under optimized conditions, aminated $TiO_2$ nanotubes (after a hydrothermal treatment at 250 ºC) result in a 4-fold increase of the water splitting current density and this effect is stable with illumination time. Combining the amination treatment with other doped methods for $TiO_2$ nanotubes or with higher quality nanostructures (higher surface area, less contaminated oxide) may result in further improvement of the photoelectrochemistry.

**Acknowledgements**

The authors would like to acknowledge the ERC, the DFG, and the DFG "Engineering of Advanced Materials" cluster of excellence for financial support. Marco Altomare (Institute for Surface Science and Corrosion LKO, University of Erlangen-Nuremberg) is acknowledged for technical help.

**Figure and table captions**

**Figure 1:** a,b and c) Top and cross-section view SEM images of bare TiO$_2$ nanotubes; d,e and f) hydrothermally treated nanotube in DETA at 250°C; g) XRD of bare TiO$_2$ nanotubes and of h) hydrothermally treated nanotubes in DETA.

**Figure 2:** a) N1s, b) C1s and c) Ti2p high resolution XPS spectra for bare TiO$_2$ nanotubes (bare NTs), hydrothermally treated NTs (at 150°C, 200°C and 250°C) and for DETA adsorbed at room temperature (adsorbed RT). d) Deconvolution of the N1s peak for the 250°C aminated TiO$_2$ nanotubes.

**Figure 3:** Selected fragments from the ToF-SIMS surface profiles of bare TiO$_2$ nanotubes (NTs) and 250°C aminated nanotubes (250°C) in positive polarity – TiO$^+$ (m/z 63.94), C$_2$H$_6$N$^+$ (m/z 44.05) and C$_3$H$_7$N$_2$O$_2^+$ (m/z 103.05); and in negative polarity – CN$^-$ (m/z 26.00), CHN$^-$ (m/z 27.00)

**Figure 4:** a) Photocurrent spectra of the TiO$_2$ nanotube before and after hydrothermal treatment at different temperature in DETA; b) Evaluation of the band gap-energy of the samples from a); c) Influence of the hydrothermal treatment time on the photocurrent spectra of the TiO$_2$ nanotube aminated at 200°C; d) Comparison of transport time constants for bare and aminated TiO$_2$ nanotube as a function of the incident photon flux, for monochromatic 396 and 475 nm light illumination.

**Figure 5:** a) Photocurrent transient vs potential curves of different hydrothermal treatment at different temperature in DETA; b) corresponding photostability experiment for 1 h at 500 mV; c) Photocurrent transient under monochromatic laser (λ = 474 nm) for the 250°C aminated nanotubes.

**Table 1:** The chemical composition (at.%) of the bare and aminated TiO$_2$ nanotubes computed from XPS data

**Figure 1**

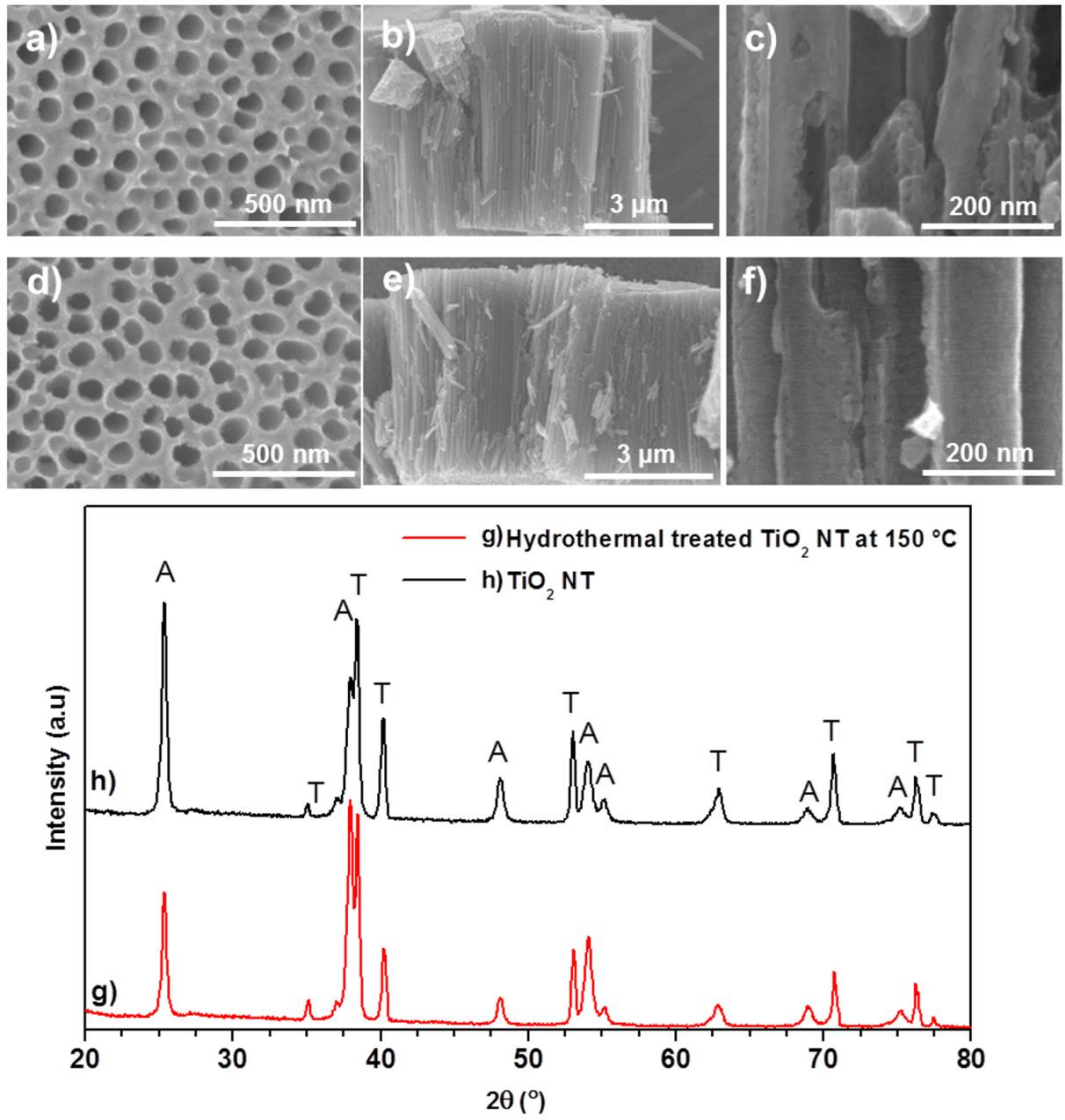

**Figure 2**

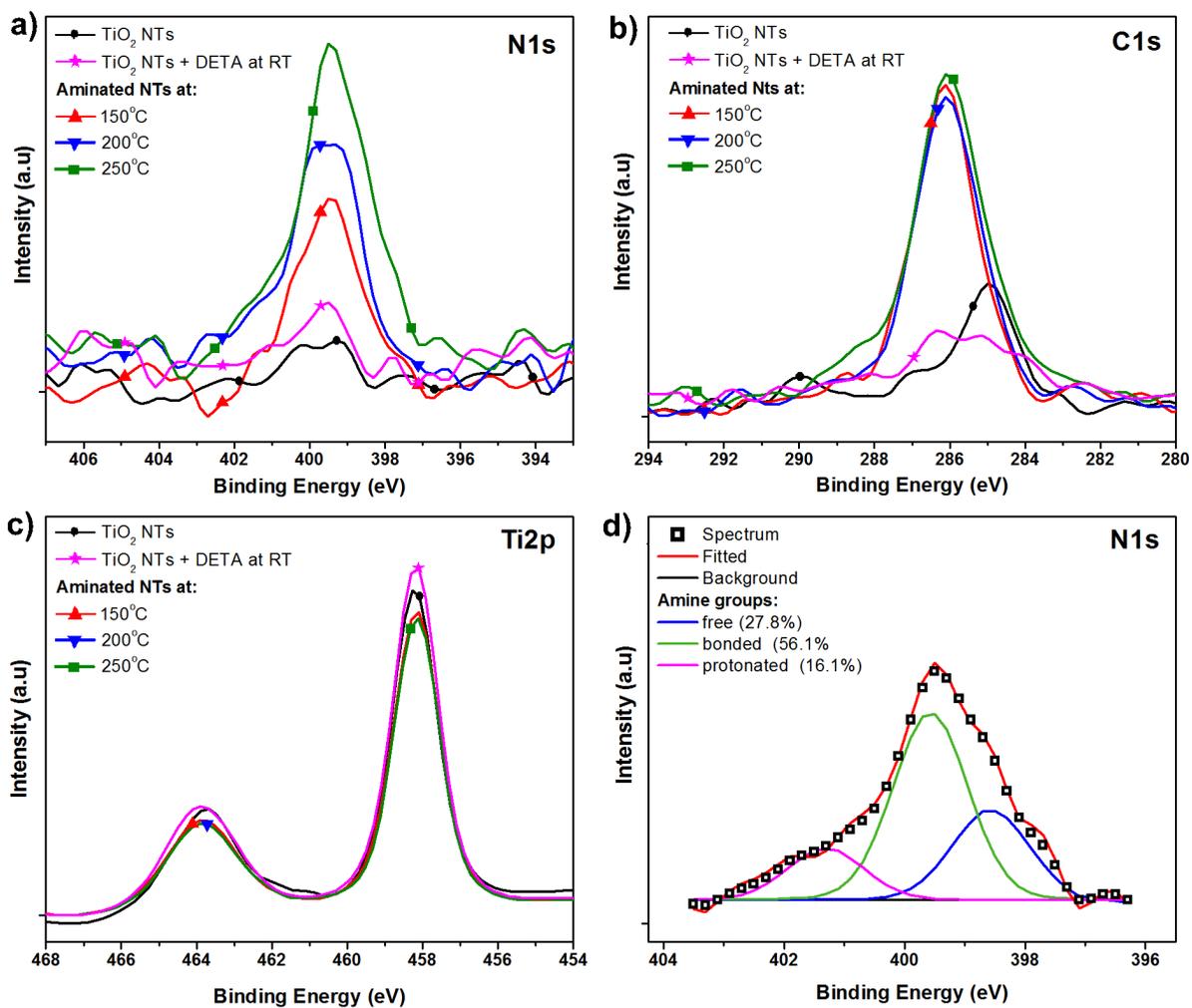

**Figure 3**

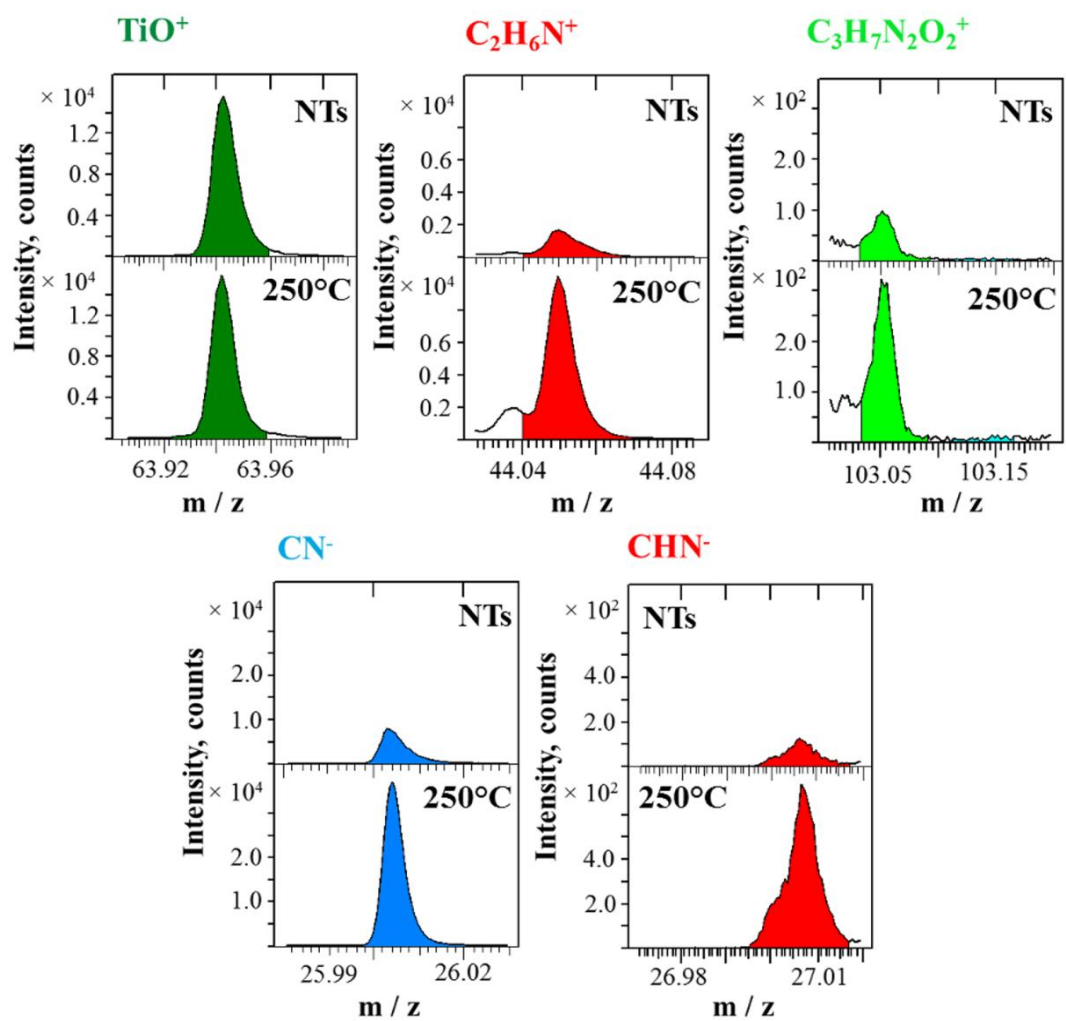

**Figure 4**

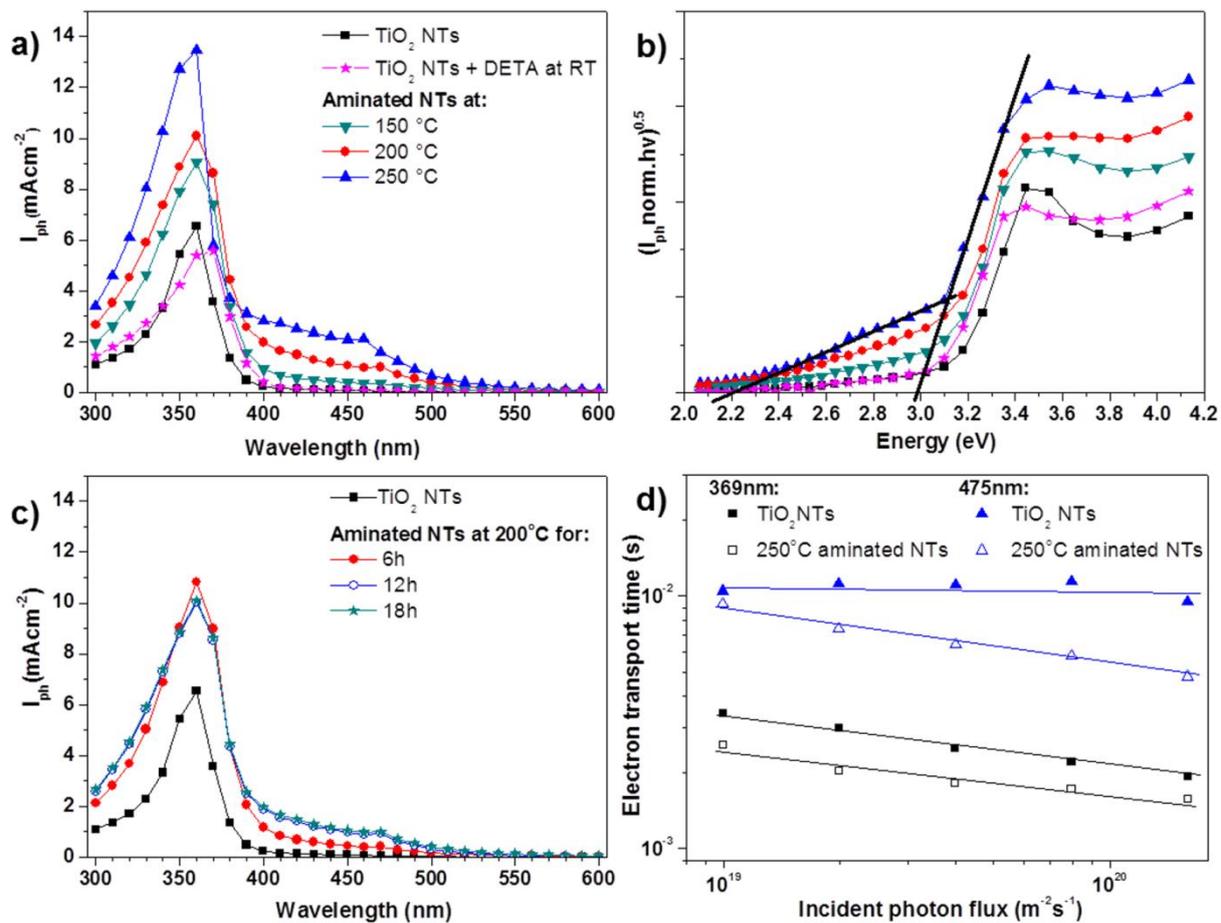

**Figure 5**

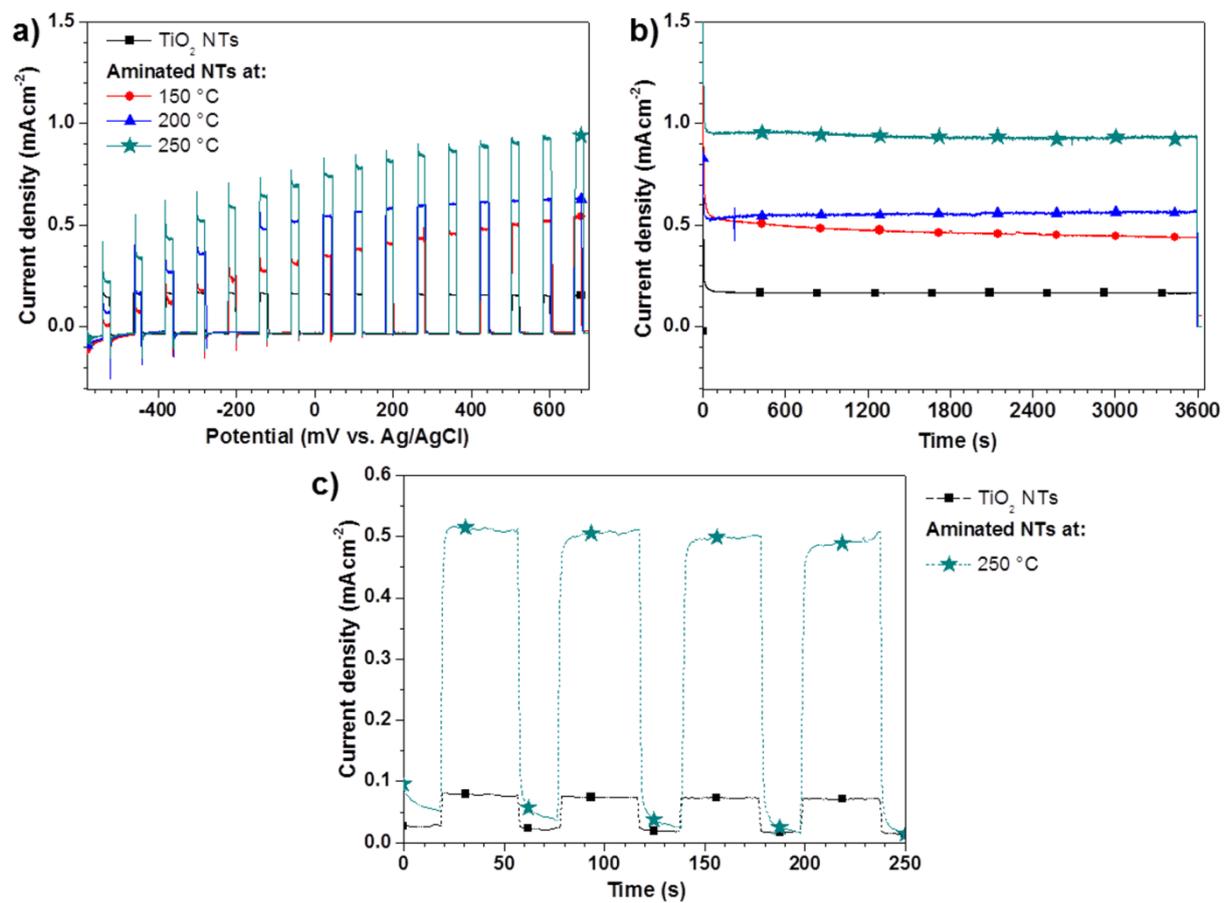

**Table 1**

| Sample | C (at.%) | N (at.%) | O (at.%) | Ti (at.%) |
|---|---|---|---|---|
| **Bare TiO$_2$ NTs** | 4.2 | 0.5 | 67.5 | 27.8 |
| **Aminated TiO$_2$ nanotubes** | | | | |
| **150°C** | 9.4 | 1.1 | 64.9 | 24.6 |
| **200°C** | 10.0 | 1.7 | 63.8 | 24.5 |
| **250°C** | 10.6 | 2.2 | 64.0 | 23.2 |
| **DETA adsorbed at room temperature (Adsorbed RT)** | 2.5 | 0.5 | 68.8 | 28.2 |
| Deconvolution of the N1s peak for aminated TiO$_2$ nanotubes | | | | |
| | Free amine, at.% | Bonded to TiO$_2$ surface, at.% | | Protonated amine, at.% |
| **150°C** | 0.62 | 0.46 | | 0.20 |
| **200°C** | 0.90 | 0.66 | | 0.14 |
| **250°C** | 0.61 | 1.24 | | 0.35 |